\documentclass{PoS}

\title{Massive Primordial Black Holes}

\ShortTitle{Massive PBH}

\author{\speaker{A.D. Dolgov}\thanks{The support of the RNF Grant 19-42-02004 is acknowledged.}\\
        Novosibirsk State University and ITEP, Russia\\
        E-mail: \email{dolgov@fe.infn.it}}

\abstract{A review  of the astronomical data of several last years on an astonishingly high amount of black holes in the 
contemporary and early ($z\sim 10$) universe  is presented. Also the data on the recently observed peculiar 
stars in the Galaxy are discussed. It is argued that practically all black holes in the universe are primordial (PBH)
and suggested that an inverted picture of the galaxy formation is realized: supermassive black holes were 
formed prior to galaxy formation and subsequently seeded the latter. Possibilities of cosmological dark matter
consisting of primordial black holes and of abundant cosmological antimatter are considered.

A mechanism of 1993  anticipating all these phenomena and predicting an extended log-normal mass spectrum
of PBH is described. }

\FullConference{Multifrequency Behaviour of High Energy Cosmic Sources - XIII - MULTIF2019\\
		3-8 June 2019\\
		Palermo, Italy}

\begin{document}

\section{Introduction \label{s-intro}}

Recent astronomical data, which keep on appearing almost every day, show that the contemporary, {${z \sim 0}$,}  
universe is much more abundantly populated by all kind of black holes, than it was expected even a few years ago. 
There are lots of black holes in all mass intervals:\\
$\bullet${massive, from a fraction of  ${M_\odot}$ up to ${\gtrsim {  100 M_\odot}}$,}\\
$\bullet${supermassive (SMBH), ${ M \sim (10^6 - 10^{9} ) M_\odot }$,}\\
$\bullet$ {intermediate mass (IMBH)
${ M\sim (10^3 - 10^5) M_\odot} $,}\\
The origin of most of these black hole is not clear, to say the least.

In addition, there is a population of peculiar stars in the Galaxy (too old, too fast, and with an unusual chemical content)
which are at odds with the conventional astrophysics.

{Moreover, the data collected during last several years indicate that  the young universe at ${z \sim 10}$ is grossly overpopulated 
with unexpectedly high amount of:} \\
{{$\bullet$} bright QSOs, alias supermassive BHs, with masses up to ${M \sim 10^{10} M_\odot}$;}\\
{$\bullet$ superluminous young galaxies;}\\
{$\bullet$ supernovae, gamma-bursters;}\\
$\bullet$ dust and heavy elements.

The canonical mechanism of  formation of SMBH even in the present day universe demands at least an order of 
magnitude longer time than the universe age,  ${t_U = 14.6\cdot 10^9}$ years,
and the problem of their formation in ten times younger universe at $z \sim 10$ is multifold more pronounced.
 
 In this talk a review of these surprising astronomical observations is presented and the mechanism
 suggested a quarter of century
 ago of primordial black hole (PBH) and peculiar star formation~\cite{AD-JS,AD-MK-NK}, 
 which neatly solves the above mentioned problems, is described. 
 Not only abundant PBHs but also a significant amount of rather strange primordial stars are predicted, 
 They may make a considerable or even 100\%  contribution to the cosmological dark matter.
 To some extent the content of this talk coincides with the author's review~\cite{AD-UFN}, but of course during
the last one-two years much more new observational data have been obtained.

\section{Black Holes: what's that \label{s-BH-types}}

\subsection{Observations of black holes \label{ss-obs-BH}}

Existence of black holes (BHs) was first envisaged by John Mitchell in 1784, who predicted that 
there might be bodies with so strong gravitational field that the second cosmic velocity would be  larger than the 
speed of light. He concluded that they neither shine not reflect light, i.e. they are absolutely dark, and it is
impossible to observe to them.

This is not precisely true, because BHs  can emit all kind of radiation in the process of their evaporation predicted by 
Stephen Hawking and, though nobody yet has seen it, there are no doubts that such process exists. The life-time of
a BH with respect to evaporation scales as its mass cube, and for $M_{BH} = 5\cdot 10^{14}$ g the life-time is close to the   
universe age $t_U \approx 10^{10} $ years.  So the stellar mass BHs (and of course heavier ones) will survive even after all 
the protons in the universe will decay, since their  life-time
is presumably finite, about $10^{33}$ years.

BHs strongly interact with the surrounding matter and it makes them very visible. For example, the
most powerful sources of radiation in the universe  are quasars (QSO) which are  pointl-like objects, within the accuracy of 
the telescopes, and shine as thousands of galaxies. The only plausible mechanism of explanation for the operation of
the central  QSO engine is the emission of radiation in the process of collisions of ultra-relativistic particles accreting to a 
supermassive black hole (SMBH) in the center. Such process can heat the surrounding matter up to million degrees.

Less massive BHs are also observed through emission of X-rays from the heated gas around them.

Another way to observe BHs is based on studying the star motion around an invisible gravitating object of small size. That's how 
the central BH in our Galaxy manifests itself.

All these methods are indirect. They  indicate only that in a small volume a large mass is concentrated. Based on the theory, 
General Relativity (GR), one may conclude that there is a BH inside this volume.

BHs may also be spotted by the gravitational lensing of some invisible objects. In this way the Massive Astrophysical Compact Halo 
Objects  (MACHOs) have been discovered. One, of course, cannot exclude that these invisible objects are e.g. some low
luminosity stars.

Because of all that some skepticism concerning an existence of BH was not eliminated, but recently it was 
strongly hit by the registration of gravitational waves (LIGO and Virgo) and  by the photo of a black hole shadow (see below).

The form of the signal created by gravitational waves is best fit to the hypothesis that it is emitted by a coalescing BH
binary with the Schwarzschild metric of each BH.
Masses of both initial BH and the final one are accurately determined and their spins are measured. 
These are the first observations proving the validity of GR for strong gravitational fields.  All numerous earlier 
tests were performed only for very weak fields.

The Event Horizon Telescope Collaboration have made a snapshot of BH shadow~\cite{BH-shadow} 
in full agreement with the GR predictions. 
The photographed  BH is situated in giant elliptical galaxy M87, at the distance 16.4 Mpc.
The measured BH mass is huge, ${ M=(6.5\pm 0.7)\cdot 10^9 M_\odot}$, It is the most
massive BH registered in the present day universe, at the redshift $z=0$. 

\subsection{GR solutions predicting black holes }

According to General Relativity four and only four different types of BHs can exist. These solutions are characterized 
by three parameters, the so called hairs, which can be measured through the gravitational and electric fields 
by an observer outside BH.  These "hairs" are the BH mass ${M}$, its electric charge ${Q}$, and the angular
momentum, spin, characterized by the dimensionless parameter ${a}$, which is confined in the limits: ${0<a<1}$.

There are four types of the (analytic!) solutions  of the General Relativity equations describing 
all four possible types of BHs:\\
$\bullet$ Schwarzschild solution (1916), the simplest BH with ${Q=a=0}$;\\
$\bullet$ Reissner-Nordstr{\"o}m  solution (1916,1918), with${Q\neq 0}$, but $a=0$; \\
$\bullet$ Kerr solutiion (1963) with ${a \neq 0}$ but $ Q=0$;\\
$\bullet$ Kerr-Newman solution (1965) $ Q\neq 0$ and $a \neq 0$.

Note that if the photon mass is nonzero, ${m_\gamma \neq 0}$,
whatever tiny, electric hairs do not exist. Coulomb field around electrically charged BH completely disappears.
There is no continuous transition from $m_\gamma \neq 0$ to $m_\gamma = 0$.

The metric around the Schwarzschild black hole with mass $M$ has the form:
\begin{equation}
ds^2 =  \eta dt^2 - (1/\eta) dr^2 - r^2 d\theta^2 - r^2  sin^2 \theta d\phi^2 ,
\label{Sch-metrci}
\end{equation}
where ${\eta} = 1- r_g/r$, with the gravitational radius (or Schwartzschildt horizon) equal to $r_g = 2M  /m_{Pl}^2 $ and 
$m_{Pl} \approx 1.2 \cdot 10^{19} $ GeV is the Planck mass. In the natural system of units (see Sec.~\ref{s-explain})
$1/m_{Pl}^2 $ is equal to the Newtonian gravitational coupling constant.

This metric has a striking property that 
an object falling into a BH never cross the horizon (by the clock of a distant observer). That's why
Zeldovich and Novikov called black holes as frozen stars. 

\subsection{Black holes by creation mechanisms \label{s-BH-prod}}

There are three possible types of BH creation and accordingly the black holes have the names: astrophysical BHs, 
supermassive BHs in galactic centers (SMBH), and primordial black holes {PBH), though such a devision is 
rather arbitrary. 

$\bullet$ {Astrophysical black holes, which are created by stellar collapse, when star exhausted its fuel and the internal pressure became 
so weak that it could not resist gravity.
Masses of such BHs are expected expected to be just above 
the neutron star masses, ${3 M_\odot}$, and normally quite close to it.
Instead the observed mass spectrum of BHs in the Galaxy has maximum at 
${M \approx 8 M_\odot}$  with the width: ${ \sim(1-2) M_\odot }$ (see below). There is no convincing explanation of
these data. 

$\bullet$ {BH crated by accretion of matter to the regions with excessive matter density.} 
{Presumably in any large galaxy there is a supermassive BH (SMBH). The masses of these black holes are in the range 
from billions solar masses, ${M\sim 10^9 M_\odot}$ in elliptic and lenticular 
galaxies and ${M\sim (10^6-10^7) M_\odot} $ in elliptic galaxies, like Milky Way.}

{However, the known mechanisms of accretion is not efficient enough to create such monsters during the 
available time equal to the universe age ${t_U \approx 15 }$ Gyr.
A longer, more than by an order of magnitude, time  duration is necessary.
{Moreover SMBH are found in very small galaxies and one SMBH lives even in almost empty space.}
{SMBH are also discovered recently in quite young universe with the age about 0.5 -1  Gyr.}
Their formation is multifold less probable.
{Massive seed are necessary, but their origin is mysterious.}

It is tempting to conclude that the SMBH are primordial and 
type of the galaxy is determined by the mass of BH and not vice versa. 

$\bullet$  {Primordial black holes (PBH) created at pre-stellar epoch in the very early universe.}
The canonical picture of their creation is the following. It might happen in the early universe that
the density contrast was accidentally very large, ${\delta\rho/\rho \sim 1}$,
at the cosmological horizon scale. Normally ${\delta\rho/\rho \sim 10^{-4} }$, but large fluctuations at small scales are
not excluded. Such a piece of volume would be inside its gravitational radius, so 
it decoupled from the cosmological expansion and became a black hole.
Such a scenario of PBH formation has been suggested by 
Zeldovich and Novikov~\cite{Z-N-pbh} and elaborated later by Carr and Hawking~\cite{C-H-pbh}.

Usually this mechanism leads to creation of PBHs with rather low masses and with sharp almost delta-function mass 
spectrum. A different mechanism suggested in our paper with J. Silk~\cite{AD-JS} (see also~\cite{AD-MK-NK})
could make PBH with masses exceeding millions solar masses and with the extended 
log-normal mass spectrum which became quite popular recently. 


\section{Puzzles of the present day universe \label{s-today}}

\subsection{Supermassive black holes today \label{ss-DSMBH-todat}}

{ Every large galaxy and even several smaller ones 
contain a central supermassive BH with mass  larger than} 
{ ${ 10^{9}M_\odot}$} in giant elliptical
and compact lenticular galaxies
and {${\sim10^6 M_\odot}$} in spiral galaxies like Milky Way.
The most massive out of them is the recently observed BH with the mass mass ${ (6.5\pm 0.7)\cdot 10^9 M_\odot}$,
see subsection~\ref{ss-obs-BH}.
{The origin of these BHs is not understood.}
{The accepted faith is that these BHs are created by matter accretion to a central seed. }
{But, the usual accretion efficiency is insufficient to create them during the Universe life-time,
14.6 Gyr.}  According to the calculations of ref.~\cite{mur} to create the
supermassive black hole SgrA* with the mass ${\sim 4\times10^6 M_\odot}$ at the 
centre of our Galaxy 
the mean accretion rate of  ${4\times 10^{-4}} M_\odot$  per year is required.
At present, X-ray observations constrain the rate of hot gas accretion 
to ${\dot{M} \sim 3 \times 10^{-6} M_\odot}$ per year 
and polarization measurements constrain it near the event horizon to 
${\dot{M}_{horizon} \sim 10^{-8} M_\odot}$/yr. 
{Thus the universe age is short by two orders of magnitude.}

{Even more puzzling: SMHBs  are observed  in small galaxies
 and even in almost {EMPTY} space, where the material to make a SMBH cannot be found.}

There are a few more pieces of evidence indicating that the conventional picture of SMBH formation is not
compatible with observations. 
{The mass of BH is typically 0.1\% of the mass of the stellar bulge of galaxy} 
 \cite{BH-bulge,BH-bulge2}
but some galaxies may  have huge BH: e.g. NGC 1277  has
the central BH  of  ${1.7 \times 10^{10} M_\odot}$, or ${60}$\% of its bulge mass.}
\cite{NGC1277}.
{This creates serious problems for the
standard scenario of the formation of central supermassive BHs by accretion of matter in the central 
part of a galaxy.}
}

Hence an inverted picture looks more plausible, when first a supermassive BH was formed and 
attracted matter being a seed for subsequent galaxy formation, as it is suggested in refs.~\cite{AD-JS,AD-MK-NK,NGC1277}.

A really striking piece of evidence in favor of creation of SMBH prior to the galaxy formation is a discovery of
"A Nearly Naked Supermassive Black Hole"\cite{BH-naked} with the host galaxy having much smaller mass than the SMBH.

There are some more systems of multiple SMBH in close vicinity, which is highly improbable in the conventional model. There are:\\
{ $\bullet$  four binaries of SMBH discovered during recent years~\cite{bin1,bin2,bin3,bin4}.}\\
{$\bullet$  a physical association of four quasars at  ${z\approx 2}$~\cite{quadr-QSO}.}
{ The probability of finding a  quadruple quasar in such close vicinity  is  ${\sim 10^{-7}}$.}\\
{$\bullet$ triple quasar~\cite{triple-QSO}.}
{Such a structure can only satisfy one of the three scenarios: a triple supermassive black hole 
(SMBH) interacting system, a triple AGN, or a recoiling SMBH.}

An orthodox point of view for the explanation of appearance of such structures is 
merging of two spiral galaxies creating an elliptical
galaxy, leaving two or more SMBHs in the center of the merged elliptical.
No other way is found in the traditional approach. However, even one SMBH is hard to create.
Heretic but simpler suggestion is that primordial SMBH formed binaries in the very early universe and 
seeded galaxy formation

\subsection{Intermediate mass black holes}

Intermediate mass black holes (MBH) are those with the masses in the interval ${M =(10^3 -10^5) M_\odot}$}.
Nobody expected them and now they came out as if from cornucopia (cornu copiae). 
According to our conjecture~\cite{AD-KP-IMBH} the
intermediate mass BHs: with ${M \sim 10^3 M_\odot}$, seeded globular clusters and IMBH with ${M\sim 10^{4}-10^5}$ 
seeded dwarf galaxies.  
{ Three years ago only 10 IMBH, were known with masses. 
${ M = 3\times10^4 - 2 \times10^5 M_\odot}$}. 
Forty of them, but somewhat more mssive, are found recently by Chandra with ${10^7<M<3\cdot 10^9}$~\cite{IMBH-Chandra} and more 
and more are appearing  in different mass intervals with an impressive rate. In ref.~\cite{IMBH-2}
the discovery of 204 black holes with masses in the range {${(1-20) \times 10^5 M_{\odot}}$."} is reported.
The group~\cite{IMBH-3} identified a sample of
{305 IMBH} candidates with ${{3\times10^4<M_{\mathrm{BH}}<2\times10^5 M_{\odot}},}$ 
residing in galaxy centers and are accreting gas that creates
characteristic signatures of a type-I active galactic nucleus.

A very interesting example of IMBH is reported in ref.~\cite{IMBH-4}. According to the authors statement
an invisible massive object in the central region of our Galaxy, based on the high-resolution molecular line observations 
with the Atacama Large Millimeter/submillimeter Array (ALMA). The morphology and kinematics of these streams can be 
reproduced well through two Keplerian orbits around a single point mass of {${(3.2 \pm 0.6) \times 10^4 M_\odot}$.} 
The results provide new circumstantial evidences for a {wandering intermediate-mass black hole in the Galactic center} 
(tramp in the galaxy), suggesting also that high-velocity compact clouds can be probes of quiescent black 
holes abound in our Galaxy.

It is tempting to conclude that such particular object is a primordial black hole, as well a all other IMBH, mentioned above, 
whose  abundance is difficult to explain other way.

Only one or two massive BH are observed in Globular clusters~\cite{BH-sh-s,bh-gc-2}. 
{Definite evidence is presented for BH with ${M\approx 2000 M_\odot}$ found
in the core of the globular cluster 47 Tucanae~\cite{BH-sh-s}.
The standard model does not have an explanation for the origin of such relatively light IMBHs. 

We assumed~\cite{AD-KP-IMBH} that such IMBHs are primordial.
If the parameters of the log-normal mass distribution of PBHs are chosen to fit the LIGO data and the density of
SMBH, then  the number of PBH with masses 
${(2-3)\times 10^3 M_\odot}$ is about ${10^4-10^5 }$ per one SMPBH with mass ${>10^4 M_\odot}$.}
{This predicted density of IMBHs is sufficient to seed the formation of all globular clusters in galaxies.}
according to the published estimates. 
{This density of IMBHs is sufficient to seed the formation of globular clusters in galaxies.}
Our prediction is that IMBHs exist  in all globular clusters but it is difficult to observe them.

 It is assumed that all PBH with 
${M > 10^4 M_\odot}$ strongly accreted matter and grew up to billion solar masses.
 
\subsection{A strange galaxy \label{ss-strange-gal}}

{Very recent a surprising discovery of
{"A young galaxy cluster in the old Universe"}~\cite{strange-gal}
was announced.  There was discovered
a blue cluster, that is a local galaxy cluster with an
unprecedentedly high fraction of blue star-forming galaxies yet hosted by a
massive dark matter halo. The blue fraction is 0.57, which is 4.0 $\sigma$
higher than those of the other comparison clusters under the same selection and
identification criteria. 
The probability to find such a high
blue fraction in an individual cluster is only 0.003\%, which challenges the
current standard frameworks of the galaxy formation and evolution in the
$\Lambda$CDM Universe. 

Such a blue cluster seems to be  in accord with our prediction of predominantly helium rich "bubbles"  in the universe. Indeed in the bubbles
with high baryon-to-photon ratio the big bang  nucleosynthesis could produce considerably higher abundance of the 
primordial helium-4, than the usual 25\%.

\subsection{Peculiar stars in the Milky Way \label{ss-unsual-stars}}

{$\bullet$ Too old stars in the Milky Way:}\\
During the last  several years the precision in determination of the stellar age has been drastically improved 
since not only uranium and thorium abundances but several other elements were used as nuclear chronometers.

Employing thorium and uranium  abundances
in comparison with each other and with several stable elements {the age of
metal-poor, halo star BD+17$^o$ 3248 was estimated as}  ${13.8\pm 4}$ Gyr~\cite{age-1}, 
which noticeably exceeds the age of inner halo of the Galaxy {${11.4\pm 0.7}$ Gyr, according
to ref.~\cite{halo-age}

The age of a star in the galactic halo, HE 1523-0901, was estimated to be 
about 13.2 Gyr~\cite{age-2}
In this work first time many different chronometers, such as the U/Th, U/Ir, Th/Eu and Th/Os ratios to
measure the star age have been employed.

The most striking example of stellar Methuselah is the 
metal deficient {high velocity} subgiant in the solar neighborhood
HD 140283  has the age ${14.46 \pm 0.31 }$ Gyr~\cite{age-3}.
{The central value of the age of this star exceeds the universe age by two standard deviations,
if, ${H= 67.3}$, as determined from the angular fluctuations of CMB and correspondingly ${t_U =13.8}$.
On the other hand, according to the recent determination of $H$ by the traditional astronomical methods, 
 ${H= 74}$, and ${ t_U = 12.5}$. So the star age exceed the universe age by more than 10 ${\sigma}$. 

This cannot be so, of course, in the conventional cosmology and astrophysics but
bur model~\cite{AD-JS,AD-MK-NK} predicts unusual  initial chemical content for a small number of stars, 
so they may look older than they are. "Our'" stars are primordial and they may have ubusually high velocity in the
Galaxy as is hinted by HD 140283. Moreover, there are more very fast stars in the Galaxy, as we see in the
following subsection.

Let us mention in conclusion a very old planet in the Kepler-10 Planetary System with the age 
${10.6^{+1.5}_{-1.3} }$ Gyr~\cite{planet-age}.
For comparison the age of the Earth is 4.54 Gyr.
The planet formation is a slow process, it should be preceded by a supernova explosion, formation of molecules, and dust.
However, as we see in what follows, the early universe is unexpectedly dusty.

{$\bullet$ High velocity and unusual chemical content stars in the Galaxy 

The Galaxy,  as discovered recently, has quite significant number of very fast stats, with velocities higher than the 
escape velocity and of stars with rather surprising chemical content. The normal stellar velocity in the galaxy is in the  
range 100-200 km/sec. Some pulsars are observed with the velocities about 1000 km/sec. The origin of the letter is
evident. Pulsars are created through the stellar collapse and even a tiny non-sphericity of the emitted neutrino flux
would create  huge recoil momentum, sufficient to accelerate the up to 1000 km/sec. However, the origin of the stars
moving faster than galactic escape velocity is puzzling.

A discovery of high speed low-mass white dwarf (LP 40-365) that 
"travels at a velocity greater than the Galactic escape velocity
and whose peculiar atmosphere is dominated by intermediate-mass elements",
is reported in ref.~\cite{hi-speed-1}
The origin of this star is mysterious. It may be a compact primordial star, predicted by the
model of refs.~\cite{AD-JS,AD-MK-NK}

Some more references to recently discovered high velocity stars in the galaxy includes the
"Old, Metal-Poor Extreme Velocity Stars in the Solar Neighborhood"~\cite{hattori}
and a special search for very such quickly moving stars:
"Gaia DR2 in 6D: Searching for the fastest stars in the Galaxy"~\cite{hi-speed-2}.
The authors conclude that the origin of the discovered stars is unclear. 
They might be accelerated to high velocity by a population of IMBH in Globular clusters, if there is sufficient number of IMBHs.

A discovery of a very unusual star is reported in ref.~\cite{hi-speed-3}.
{The authors found red host star and planet masses of 
$M_{\rm host} =0.15^{+0.27}_{-0.10}M_\odot$} and $m_p=18^{+34}_{-12}M_\oplus$.
{The life-time of main sequence star with the solar chemical content is larger than $t_U$ already for 
$ M< 0.8 M_\odot  $.}
The origin is unknown. It may be a primordial helium  star? 

Another puzzling discovery of
"A class of partly burnt runaway stellar remnants from peculiar
  thermonuclear supernovae" is announced in~\cite{hi-speed-4}.
 { Discovery of three  chemically peculiar runaway stars,
 survivors of thermonuclear explosions - according to the authors. }
"With masses and radii ranging between 0.20-0.28 $M_\odot$ and 0.16-0.60 $R_\odot$,
respectively, as is stated in the paper, these {inflated white dwarfs} are the partly burnt
remnants of either peculiar Type SNIa or electron-capture supernovae".

And at last, the discovery of this year~\cite{hi-speed-4}:
{"A hyper-runaway white dwarf in Gaia DR2 as a Type Iax supernova primary
  remnant candidate"}
 Quoting the authors: "We report the likely first known example of an unbound white dwarf that is
consistent with being the fully-cooled primary remnant to a Type Iax supernova.
The candidate, LP 93-21, is travelling with a galactocentric velocity of
v$_{gal}$ $\simeq$ 605 km s$^{-1}$, and is gravitationally unbound to the Milky
Way, We rule out an extragalactic origin. The Type Iax supernova ejection
scenario is consistent with its peculiar unbound trajectory, given anomalous
elemental abundances are detected in its photosphere via spectroscopic
follow-up. This discovery reflects recent models that suggest stellar ejections
likely occur often." This could be a peculiar WD or a remnant of a primordial star.

\subsection{MACHOs \label{ss-macho}}

MACHOs is the abbreviation for Massive Astrophysical Compact Halo Objects 
discovered through gravitational microlensing by Macho and Eros groups.
They are invisible (very weakly luminous or even non-luminous) objects}
masses about a half of the solar mass
in the Galactic halo, in the center of the Galaxy, and recently in
the Andromeda (M31) galaxy. 
Their density is  significantly greater than the density expected from the known low luminosity
stars and the BH of similar mass. 

The data on the MACHO abundance presented by different groups  are rather controversial. 
The present date situation is reviewed and summarized in 
refs.~\cite{moniez,sib-ufn,Blinnikov:2014nea,bdpp,AD-SP}. Briefly, the situation is the following. 

MACHO group \cite{MACHO2000} reported registration of 13 - 17 microlensing events  towards the Large Magellanic Cloud (LMC),
which is significantly higher than the number which could originate from the known low luminosity stars. On the other hand  this amount
is not sufficient to explain all dark matter in the halo. The fraction of the mass density of the observed objects, which created the
microlensing effects, with respect to the energy density of the dark matter in the galactic halo, $f$, according to the 
observations~\cite{MACHO2000} is in the interval:
\begin{equation}
0.08<f<0.50 ,
\label{f-macho}
\end{equation}
at 95\% CL for the mass range $  0.15M_\odot < M < 0.9M_\odot  $.

EROS collaboration~\cite{EROS-1}  has placed the upper
limit on the halo fraction, $f<0.2$ (95\% CL) for { the} objects in the specified above MACHO
mass range, while EROS-2 \cite{Tisserand:2006zx} gives $ f<0.1$ for $0.6 \times 10^{-7}M_\odot<M<15 M_\odot$
for the survey of Large Magellanic Clouds.
It is considerably less than that measured by the MACHO collaboration in the central region of the LMC.

The new  analysis of 2013 
by EROS-2, OGLE-II, and OGLE-III collaborations~\cite{Novati:2013fxa} towards the Small Magellanic Cloud (SMC). 
revealed five  microlensing events towards the SMC (one by EROS and four by OGLE), which lead to the upper limits
 $ {f <0.1} $ obtained at 95\% confidence level for MACHO's with the mass $ 10^{-2} M_\odot$
and $ {f <0.2} $ for MACHOs with the mass $ 0.5 M_\odot$.

Search for microlensing in the direction of Andromeda galaxy (M31) demonstrated some contradicting 
results~\cite{moniez,sib-ufn} with an uncertain conclusion. E.g. AGAPE collaboration \cite{AGAPE2008}, 
finds the halo MACHO fraction in the range $0.2<f<0.9$.
while MEGA group presented the upper limit $f<0.3$~\cite{MEGA2007}.
On the other hand, the recent discovery of  10 new microlensing events~\cite{Lee-2015}  
is very much in favor of MACHO existence. The authors conclude: ``statistical studies and individual microlensing events
point to a non-negligible MACHO population, though the fraction in the halo mass remains uncertain''. 

Some more recent observational data and  the other aspects  of the microlensing are discussed in ref.~\cite{Mao2012}.

It would be exciting if all DM were constituted by old stars and black holes
made from the high density baryon bubbles 
as suggested in refs.~\cite{AD-JS,AD-MK-NK} 
with masses in still allowed
intervals,  but more detailed analysis of this possibility has to be done.

There is a series of papers claiming the end of MACHO era. For example in ref.~\cite{YooCG2004} the authors 
stated "we exclude MACHOs with masses $M > 43 M_\odot $ at the standard local halo density. This removes 
the last permitted window for a full MACHO halo for masses $ M> 10^{-7.5} M_\odot$.

In addition to the criticism raised in the paper~\cite{YooCG2004}, some more arguments against abundant
galactic population of MACHOs  are also presented in ref.~\cite{BEDu-2003,BEDu-2004}.
However, according to the paper~\cite{griest},  the approach of the mentioned works have serious
flaws and so their results are questionable. A reply to this criticism is presented in the subsequent 
paper~\cite{EvansBel2007}. 

The data in support  of smaller density of MACHOs in the direction to SMC
is presented in ref.~\cite{Tisserand:2006zx} 

Later, however, another paper of the  Cambridge group \cite{Quinn:2009zg} was published where, on the basis of
studies of binary stars, arguments in favor of real existence of MACHOs and against the pessimistic
conclusions of ref.~\cite{YooCG2004} were presented.

The latest investigation on the "end of MACHO era" was presented in ref.~\cite{Monroy-2014}, where it is concluded 
that "the upper bound of the MACHO mass tends to less than $5 M_\odot$
does not differ much from the previous one. Together with microlensing studies that provide lower limits on the MACHO mass, 
our results essentially exclude the existence of such objects in the galactic halo".

A nice review of the state of the art and some new data is
presented in ref.~\cite{Lee-2015} with the conclusion that
some statistical studies and individual microlensing events point to a non-negligible MACHO population, though the fraction 
in the halo mass remains uncertain.

According to the results of different groups the fraction of MACHO mass density with respect to the total
mass density of dark matter varies in rather wide range:
\begin{equation}
f_{MACHO} = \frac{\rho_{MACHO}}{\rho_{DM}} \sim (0.01 - 0.1)
\label{f-MACHO}
\end{equation}
Notice a large variance of the results by different groups. 

An interesting option is that the spatial distribution of  MACHOs may be very inhomogeneous and non-isotropic. 
Due to selection effect, MACHOs are observed
only in over-dense clumps  where their density is much higher than the average one. For a review and
the list of references on dark matter clumping see e.g.~\cite{DM-clump}. Clumping of primordial back holes, 
due to dynamical friction, may be much
stronger than the clumping of dark matter consisting of elementary particles.
This hypothesis would allow to avoid contradiction between the observed high density of MACHOs and the predicted  much smaller density of them based on the log-normal mass spectrum with  $M_m = (7-9) M_\odot $.

\subsection{Mass spectrum of astrophysical  BH in the Galaxy \label{ss-mass-spectrum-gal}} 

It was discovered during the last decade that the BH masses in the Galaxy are concentrated in the narrow range 
${ (7.8 \pm 1.2) M_\odot }$~\cite{bh-gal1}. 
This result agrees with another paper where
a peak around ${8M_\odot}$, a paucity of sources with masses below
 ${5M_\odot}$, and a sharp drop-off above
${10M_\odot}$ are observed~\cite{bh-gal2}, 
{These features are not easily explained  in the standard model of BH
formation by stellar collapse.}

\subsection{Gravitational waves from BH binaries \label{ss-GW-Bh-bin}}

Registration of gravitational waves (GW) from coalescing  BH binaries solved long existed problems
of test of GR for strong field and presented a first direct proof of BH existence. However, it opens
immediately several more problems. "In much wisdom is much grief".  In short there are three
very interesting problems addressed in our paper~\cite{bdpp}:\\
$\bullet$ {1. Origin of heavy BHs  with the masses ${\sim 30 M_\odot}$.}\\
$\bullet$  {2. Formation of BH binaries from the original stellar binaries. }\\
$\bullet$ {3.  Low spins of the coalescing BHs .}\\
{1.  {\it Massive BH origin.}  Such BHs are believed
to be created by massive star collapse,} though a convincing theory is still lacking.
{To form so heavy BHs, the progenitors should have ${M > 100 M_\odot}$
i.e. the  masses two orders of magnitude higher than the solar mass 
and  a low metal abundance to avoid too much
mass loss during the evolution.} Such heavy stars might be present in
young star-forming galaxies {but they are not observed in the necessary amount.}

Mass of BH in LIGO binaries is up to $~50M_\odot$  (surprise, but can be reconciled with stellar evolution), rumors 
that a $100 M_\odot$ was found  - a puzzle for stellar evolution.
(borrowed from K.Postnov talk at Isaak Markovich Khalatnikov  Centennial Conference).

{On the other hand, primordial BH with the observed by LIGO masses may be  created with sufficiently high density. }

2. {\it Formation of BH binaries.} Stellar binaries were 
formed from common interstellar gas clouds and are quite frequent in galaxies.
{If BH is created through stellar collapse,} {a small non-sphericity leads to a very large
recoil velocity of the BH and the binary is destroyed.} 
{An example of such effect is  demonstrated by huge velocities of pulsars in the Galaxy.)}
{BH formation from PopIII stars and subsequent formation of BH
binaries with 
${(36+29) M_\odot}$ is analyzed and 
found to be negligible. } 

{The problem of the binary formation is simply solved if the observed sources of GWs are the binaries of
primordial black holes (PBH).} 
{They were at rest in the comoving volume and may have non-negligible probability
to become gravitationally bound.}

3.  {\it The low value of the BH spins in GW150914 and in almost all (except for three) other events.}
It strongly constrains astrophysical BH formation from close binary systems. 
{Astrophysical BH are expected to have considerable angular momentum but  still the 
dynamical formation of double massive low-spin BHs in dense stellar clusters is not excluded, though difficult.} 
{On the other hand, PBH practically do  not rotate because vorticity perturbations 
in the early universe are known to be vanishingly small.}

{However, individual PBH forming a binary initially rotating on elliptic orbit could gain collinear spins about 0.1 - 0.3,
rising with the PBH masses and eccentricity~\cite{post-mit,PKM}
 This result is in agreement with the GW170729 LIGO event produced by the binary with masses ${50 M_\odot}$  and ${30 M_\odot}$ 
 and probably with GW151216 .} 
{Earlier  in the works~\cite{mirb,de-luca} 
much weaker gain of the angular momentum gain was obtained.}

\section{Surprises of the Early Universe \label{s-early-universe}}

{The data collected during last several years indicate that  the young universe at ${z \sim 10}$ is grossly overpopulated 
with unexpectedly high amount of:} 

\subsection{Superluminous young galaxies \label{ss-super-gal}}

Several galaxies have been observed at high redshifts,
with natural gravitational lens ``telescopes,}
e.g. a galaxy at {${z \approx 9.6}$} which was created when the universe was about
0.5 Gyr old~\cite{gal-9.6} 
Moreover, galaxy at {${z \approx 11}$} has been detected 
which was formed earlier than the universe age was { 0.41 Gyr} (or even shorter with larger H).
This galaxy is three times more luminous in UV than other galaxies at ${z = 6-8}$. 
The discovery of another unexpectedly early created galaxy at red-shift  $z=11$ is reported in ref.~\cite{gal-11-2}.

The discovery of not so young but extremely luminous galaxy with the luminosity {${L= 3\cdot 10^{14} L_\odot }$   
was announced in ref.~\cite{gal-Tsai}.This galaxy already existed, 
when the universe age was ${t_U \sim 1.3 }$ Gyr.
{The galactic seeds, or embryonic black holes, necessary for the creatkon of such a huge galaxy might be bigger than thought 
possible.} Ona of the authors of the work P. Eisenhardt: said "How do you get an elephant?  One way is start with a baby elephant."
{The BH was already billions of ${M_\odot}$, when our universe was only a 
tenth of its present age of 13.8 billion years.}
"Another way to grow this big is to have gone on a sustained binge, consuming food 
faster than typically thought possible."   {A low spin of the seed is necessary.}

To finish with the list of these surprises let us mention almost yesterday discovery, reported in
ref.~\cite{wang-gal}, of  the submillimeter (wavelength 870um) detections of
39 massive star-forming galaxies at $z > 3$, which are unseen in the spectral
region from the deepest ultraviolet to the near-infrared. 
They contribute a tota star-formation-rate density ten times larger than that of equivalently massive
ultraviolet-bright galaxies at $z >3$. Residing in the most massive dark matter
halos at their redshifts, they are probably the progenitors of the largest
present-day galaxies in massive groups and clusters. 
{Such a high abundance of
massive and dusty galaxies in the early universe challenges our understanding
of massive-galaxy formation.}

An onset of the star formation 250 million years after the Big Bang is  claimed in ref.~\cite{stars-early}.
The authors observed the oxygen line at ${z = 9.1096\pm0.0006}$. According to their conclusion
"This precisely determined redshift indicates that the red rest-frame optical colour arises from a dominant stellar component that 
formed about 200 million years after the Big Bang, corresponding to a redshift of about 15.
Although we are observing a secondary episode of star formation at ${z = 9.1}$, the galaxy formed the bulk of its stars 
at a much earlier epoch." 

As is stated in the paper "Monsters in the Dark"~\cite{monsters}, 
{the density of galaxies at ${z \approx 11}$ is 
${10^{-6} }$ Mpc${^{-3}}$, an order of magnitude higher than estimated from the data at lower z.}
{The origin of these galaxies is unclear.}

According to ref.~\cite{melia} 
{"Rapid emergence of high-z galaxies so soon after big bang} 
may actually be in conflict with current understanding of how they came to be. This
problem is very reminiscent of the better known (and
probably related) {premature appearance of supermassive black holes at ${z\sim 6}$.} It is difficult to understand how
{${10^9 M_\odot}$} black holes appeared so quickly after the big
bang {without invoking non-standard accretion physics and the formation of massive seeds,} 
{both of which are not seen in the local Universe."}

\subsection{Supermassive BH and/or QSO \label{ss-SMBH-early}}

{Another and even more striking example of early formed objects are high z quasars.}
About 40 quasars with $z> 6$ were known three years ago, each quasar containing BH with 
${M \sim 10^9 M_\odot}$. \\
The maximum redshift  QSO is discovered in ref.~\cite{QSO-7.085}
 at the redshift {${ z = 7.085}$}, luminocity {${L \approx 6 \cdot 10^{13} L_\odot}$,}
and the mass {${M=2 \cdot 10^9 M_\odot}$,} 
The quasar was  formed before the universe reached {${0.75}$ Gyr. } 

In addition to this forty quasars, one more monster was discovered~\cite{qso-max}
{with the outrageously huge mass ${ 1.2\cdot 10^{10} M_\odot}$.
{There is already a serious problem with formation of lighter and less luminous quasars}
{which is multifold deepened with this new "creature".}
{The new one with more than 10 billion solas masses is absolutely forbidden
in the standard approach.} 

Recent observations by SUBARU practically doubled the number of discovered high z quasars~\cite{matsuoka}
In particular this group reported the first  low luminosity QSO at $ z >7$.

The accretion rate, as estimated in ref.~\cite{early-accret} is about 2200 ${M_{\odot}}$ during 320 Myr in a
halo with a mass of ${ 3 \times 10^{10}~M_{\odot}}$ at ${z=7.5}$. It is by far less than is necessary.

The observation of a 800 million solar mass black hole in a significantly neutral universe at redshift 7.5~\cite{neutral}
 indicated that accretion in this particular case is absent, otherwise the plasma should be ionized.

It is difficult to understand how {${10^9 M_\odot}$} black holes  {(to say nothing about ${10^{10} M_\odot}$)
appeared so quickly after the big bang {without invoking non-standard accretion physics
and the formation of massive seeds,
{both of which are not seen in the local Universe.}

To conclude on QSO/SMBH:\\
The quasars are supposed to be supermassive black holes}
{and their formation in such short time by conventional mechanisms looks problematic, to say the least.} \\
{Such black holes, formed when the Universe was less than one billion years old,} 
present substantial challenges to theories of the formation and growth of
black holes and the coevolution of black holes and galaxies.}\\
Even the origin of SMBH in contemporary universe during 14 Gyr is difficult to explain. 
{Non-standard accretion physics and the formation of massive seeds seem to be necessary.
Neither of them is observed in the present day universe.}

\subsection{Evolved chemistry, dust, supernovae, and gamma-bursters
\label{ss-sn-gamma}}

{The medium around the observed early quasars contains
considerable amount of ``metals''} (elements heavier than He). 
According to the standard picture, only elements up to ${^4}$He  { and traces of Li, Be, B}
were formed by big bang hucleosynthesis (BBN), which took place when the universe was about a 100 second
old. The heavier elements, according to the conventional cosmology, were created much later
by stellar nucleosynthesis and} {dispersed in the interstellar space by supernova explosions.}
{Hence, an evident but not necessarily true conclusion was
that prior to or simultaneously with the QSO formation a rapid star formation should take place.}
{These stars should evolve to a large number of
supernovae enriching interstellar space by metals through their explosions.}

Another possibility is a non-standard BBN in bubbles with very high baryonic density, which allows for 
formation of heavy elements beyond lithium  in the very early universe,
as suggested in refs~\cite{AD-JS,AD-MK-NK}.

The universe at ${z >6}$ is  not only enriched by metals but also is quite dusty, as observed in the
papers~\cite{dust-1,dust-2}. Abundant dust is observed in several early  galaxies, e.g. in HFLS3 at ${ z=6.34} $ and in
A1689-zD1 at ${ z = 7.55}$~\cite{dust-cnfrm}.} 
{Copious Amounts of Dust and Gas is seen in a z=7.5 Quasar Host Galaxy~\cite{dust-3}.
{Dusty galaxies show up at redshifts corresponding to a Universe which is only about 500 Myr old.}
Very high past star formation is needed to explain the presence of $ \sim10^8 M_\odot$ of dust implied by the 
observations~\cite{mancini}.
.

Later made catalogue of the observed dusty sources~\cite{dust-catalogue} confirms earlier observations and 
indicates that the number of dusty sources is an order of magnitude larger than  predicted by the canonical theory.

An analysis of the observations and a review of possible scenarios of  dust dust production  
in galaxies at ${z \sim  6-8.3}$ is presented in~\cite{dust-rev}. The authors conclude that
 {the mechanism of dust formation in galaxies at high redshift is still
unknown.} Asymptotic giant branch (AGB) stars and explosions of supernovae (SNe)
are possible dust producers, and non-stellar processes may substantially
contribute to dust production. However, AGB stars are not efficient enough
to produce the amounts of dust observed in the galaxies. 

{In order to explain these dust masses, SNe would have to have maximum efficiency and not
destroy the dust which they formed.} Therefore, the observed amounts of dust in
the galaxies in the early universe were formed either by efficient supernovae
{or by a non-stellar mechanism, for instance the grain growth in the
interstellar medium.}

Or non-standard big bang nucleosynthesis with large baryon-to-$\gamma$ ratio leading to abundant formation of 
heavy elements, see below.

{To make dust a long succession of processes is necessary:} 
{first, supernovae  explode to deliver 
heavy elements into space (metals),} 
{then metals cool and form molecules,}
{and lastly molecules make dust which could form macroscopic
pieces of matter,}  turning subsequently into early rocky planets.  Note in brackets that
we all are dust from SN explosions, at much later time. 

Existence of heavy elements, molecules, and dust in the early universe indicates 
but there already could be life .Several hundred million years may be enough for birth of living creatures. 

{Observations of high redshift gamma ray bursters (GBR) also indicate 
a high abundance of supernova at large redshifts.} 
{The highest redshift of the observed GBR is 9.4 and there are a few more
GBRs with smaller but still high redshifts.}

{Dust production scenarios in galaxies at high ${z=6-8.3 }$ was reviewed in ref.~\cite{dust-rev}. 
 {The authors concluded that the  mechanism of dust formation in galaxies at high redshift is still
unknown.} Asymptotic giant branch (AGB) stars and explosions of supernovae (SNe)
are possible dust producers, and non-stellar processes may substantially
contribute to dust production. However, AGS are not efficient enough
to produce the amounts of dust observed in the galaxies. \\
{In order to explain these dust masses, SNe would have to have maximum efficiency and not
destroy the dust which they formed.} Therefore, the observed amounts of dust in
the galaxies in the early universe were formed either by efficient supernovae
{or by a non-stellar mechanism, for instance the grain growth in the
interstellar medium.}

{Observations of high redshift gamma ray bursters (GBR) also indicate 
a high abundance of supernova at large redshifts~\cite{GRB-gen}.} 
The highest redshift of the observed GBR is 9.4 and there are a few more
GBRs with smaller but still high redshifts.

{The necessary star formation rate for explanation of these early
GBRs is at odds with the canonical star formation theory.}
Non-standard big bang nucleosynthesis with large baryon-to-photon ratio leading to formation of 
heavy elements in the very early universe may be  needed.

These facts are in good agreement with the our 1993 model~\cite{AD-JS}
but in tension with the Standard Cosmological Model.

\section{Possible mechanism explaining the surprises in the data \label{s-explain}}

In 1993 we suggested a model~{AD-JS} of PBH creation which allowed an abundant 
formation of very massive black holes with the
log-normal mass spectrum:
\begin{equation}
\frac{dN}{dM} = \mu^2 \exp{[-\gamma \ln^2 (M/M_0)], }
\label{dn-dM}
\end{equation}
depending only on three constant parameters: ${\mu}$, ${\gamma}$,  ${M_0}$. 
{It can be easily generalized to a superposition of several log-normal spectra with
several separated maxima.}

Here we use the natural system of units where the speed of light, the reduced Planck constant, and the Boltzmann constant 
are all equal to unity:$c=h/(2\pi) = k =1$. In this system the Newtonian gravitational constant is
$G_N = 1/ m_{Pl}^2$, where th Planck mass is $m_{Pl} = 1.22\cdot 10^{19} $ GeV.

{Such a simple  form pf the PBH mass spectrum is a result result of  quantum diffusion of baryonic scalar field during inflation. 
Probably such spectrum is a general consequence of diffusion.}

Log-normal mass spectrum of PBHs was rediscovered later in 
ref.~\cite{clesse}. Now in many works such spectrum  is postulated without any justification.
In our model the PBH creation was based on the supersymmetry (SUSY) motivated baryogenesis, 
proposed by Affleck and Dine (AD)~\cite{AD}.
SUSY predicts existence of  scalars with  non-zero baryon number, {${ B\neq 0}$.}
Such scalar bosons may condense along {flat} directions of the quartic potential:
\begin{equation}
U_\lambda(\chi) = \lambda |\chi|^4 \left( 1- \cos 4\theta \right)
\end{equation}
and/or of the mass term, ${ m^2 \chi^2 + m^{*\,2}\chi^{*\,2}}$:
\begin{equation}
U_m( \chi ) = m^2 |\chi|^2 \left[{ 1-\cos (2\theta+2\alpha)}  \right],
\end{equation}
where ${ \chi = |\chi| \exp (i\theta)}$ and ${ m=|m|e^\alpha}$.
{If ${\alpha \neq 0}$, charge symmetry (between particles and antiparticles) C and CP are  broken.}

In Grand Unified version of SUSY baryonic number is naturally non-conserved, which is reflected in non-invariance of ${U(\chi)}$
w.r.t. phase rotation.

{ Initially (after inflation) ${\chi}$ is normally away from the origin and, when 
inflation is over, starts to evolve down to the equilibrium point, ${\chi =0}$,
Its evolution is described by the equation similar to that governing the motin of a point-like
particle  in Newtonian mechanics:}
\begin{equation}
\ddot \chi +3H\dot \chi +U' (\chi) = 0.
\label{ddot-chi}
\end{equation}
Baryonic charge of $ \chi$ is equal to :
\begin{equation}
B_\chi =\dot\theta |\chi|^2.
\end{equation}
Due to instability of effectively massless field in De Sitter like stage $\chi$ can acquire a large baryonic namber.
It is analogous to mechanical angular momentum in two dimensional complex $\chi$-plane. 
The ${{\chi}}$ decays transferred the accumulated baryonic charge to that of quarks in B-conserving process.
 {AD baryogenesis could lead to baryon asymmetry of order of unity, much larger
than the observed  ${10^{-9}}$.}

If $ { m\neq 0}$, the angular momentum, B, can generated by a different 
direction of the  quartic and quadratic valleys at low ${\chi}$.
{If CP-odd phase ${\alpha}$ is small but non-vanishing, both baryonic and 
antibaryonic domains might be  formed}
{with possible dominance of one of them.}\\
{Matter and antimatter domains may exist but globally ${ B\neq 0}$.}
 
We slightly generalized the original Affleck-Dine model introducing a general renormalizable coupling of $\chi$
to the inlfaton field $\Phi$ (the first term in the equation below):
\begin{equation}
U = {g|\chi|^2 (\Phi -\Phi_1)^2}  +
\lambda |\chi|^4 \,\ln ( \frac{|\chi|^2 }{\sigma^2 } )
+\lambda_1 (\chi^4 + h.c. ) + 
(m^2 \chi^2 + h.c.). 
\end{equation}
The other terms in this potential presents the usual Affleck-Dine potential with one loop radiation corrections.
We assume that the inflaton $\Phi$ passes through the value $\Phi_1$ still during inflation but not too close to
the end of it.

If $\Phi$ is close to $\Phi_1$ the window in the potential to the flat  directions is open and $\chi$ can quantum-fluctuate 
to acquire rather high amplitude. It evolves according to quantum diffusion
equation derived by Starobinsky, generalized to a complex field ${\chi}$.

If the window to flat direction, when ${\Phi \approx \Phi_1}$ is open only {during 
a short period,} cosmologically small but possibly astronomically large 
bubbles with high ${ \beta}$ could be created, occupying {a small
fraction of the universe,} while the rest of the universe has normal
{${{ \beta \approx 6\cdot 10^{-10}}}$, created 
by small ${\chi}$}. 

Initially the bubble with large $\beta$ would have have the almost the same density as the average cosmological
density because before the QCD phase transition quarks are essentially massless and the initial perturbations are 
the isocurvature ones, though they might have quite large amplitude. Only after the QCD phase transition when  massless 
quarks turn into heavy nucleons the original isocurvature perturbations transformed into density perturbations. The bubbles 
with high baryonic number can either form PBH or compact stellar like objects, depending on their magnitude and size.

Such mechanism of massive PBH formation is very much different from other ones studied in the literature.
{The fundament of PBH creation is build at inflation by making large isocurvature
fluctuations at relatively small scales, with practically vanishing density perturbations.} 
{Initial isocurvature perturbations are in chemical content of massless quarks.
Density perturbations are generated rather late after the QCD phase transition.}

{The emerging universe looks like a piece of Swiss cheese, where holes are high baryonic 
density objects occupying a minor fraction of the universe volume.}

{The outcome, depending on ${\beta = n_B/n_\gamma}$, could be:}\\
$\bullet$ PBHs with log-normal mass spectrum.\\
$\bullet$ {Compact stellar-like objects, as e.g. cores of red giants.}\\
$\bullet${Disperse hydrogen and helium clouds  with (much) higher than average ${n_B}$ density.}\\
$\bullet$ {${\beta}$ may be negative leading to compact antistars which could survive annihilation with the 
homogeneous baryonic background.}

A modification of inflaton interaction with scalar baryons as e.g.
\begin{equation} 
U \sim |\chi|^2 (\Phi - \Phi_1)^2 ((\Phi - \Phi_2)^2
\end{equation}
gives rise to a superposition of two log-normal spectra or multi-log.

\section{Conclusion \label{conclude}}
\noindent
$\bullet${ 1. Natural baryogenesis model leads to abundant formation of PBHs and compact stellar-like
objects in the early universe after QCD phase transition, ${t \gtrsim 10^{-5}} $ sec.}\\
$\bullet${ 2. These objects had originally log-normal mass spectrum, though it may be subsequently distorted
due to matter accretion, especially for very heavy ones.} \\
$\bullet$ 3. PBHs formed at this scenario can explain the peculiar features of the sources
of GWs observed by LIGO.\\
 $\bullet$ 4. The considered mechanism solves the numerous mysteries of ${z \sim 10}$ universe: abundant population
of supermassive black holes, early created gamma-bursters and supernovae, early bright galaxies, and evolved chemistry including dust.\\
$\bullet$ {5. There is persuasive data in favor of the inverted picture of galaxy formation, when first a supermassive 
BH seeds  are formed and later they accrete  matter forming galaxies.}\\
$\bullet$ 6. Inverted picture of galaxy formation is advocated.\\
$\bullet${ 7. An existence of supermassive black holes observed  in all large and some small galaxies and even in
almost empty environment is naturally explained.}\\
$\bullet$ {8. "Older than ${t_U}$" stars may exist; the older age is mimicked by the unusual initial chemistry. }\\
$\bullet$ {9. Existence of high density invisible "stars" (machos) is understood.} \\
$\bullet$ 10. Explanation of origin of BHs with 2000 ${M_\odot}$ in the core of globular cluster and the
observed density of GCs is presented.\\
$\bullet$ {11. A large number of the recently observed IMBH was  predicted.}\\
$\bullet${12. A large fraction of dark matter or 100\% can be made of PBHs.}\\
{ {13. Clouds of matter with high baryon-to-photon ratio may exist.}}\\
$\bullet$ {14. A possible by-product: plenty of (compact) anti-stars, even in the Galaxy,}
{not yet excluded by observations.}

Extreme claims:\\
{$\bullet$ {Black holes in the universe are mostly primordial.}}\\
{$\bullet$ {Primordial BHs make all or dominant part of dark matter.}}\\
{$\bullet$ {All QSO were created in the very early universe.}}\\
 $\bullet$ Metals and dust are made much earlier than at ${z = 10}$.\\
{$\bullet$ {Inverted picture of galaxy formation: seeding of galaxies by SMPBH  or IMPBH;}}\\
 $\bullet$ Seeding of globular clusters by ${10^3 - 10^4}$ BHs,
{dwarfs by ${10^4 - 10^5}$ BH.}\\

\bigskip
\bigskip
\noindent {\bf DISCUSSION}

\bigskip
\noindent {\bf BATISTELLI:} 
Do you foresee primordial black holes as a source of spectral distiortion of CMBR?

\bigskip
\noindent {\bf DOLGOV:} Yes, they can create spectral distortion at the level which leads to noticeable constraints in some
interesting mass intervals.

\end{document}